\documentclass[12pt]{article}
\usepackage{sproclmo}
\usepackage{epsfig}
 \newcommand \be {\begin{equation}}
\newcommand \ee {\end{equation}}
 \newcommand \bea {\begin{eqnarray}}
\newcommand \eea {\end{eqnarray}}
\newcommand \nn \nonumber
\def \(({\left(}
\def \)){\right)}

\begin{document}
 \bibliographystyle{unsrt}
 \begin{flushright}
 { \rm LPTENS  /02/08    }
\end{flushright}

\title{ HARTREE-FOCK  MODEL  OF A SELF-AVOIDING   FLEXIBLE  POLYMER}

\author{C. BOUCHIAT }
 \address{Laboratoire de Physique Th\'eorique de l'Ecole Normale Sup\'erieure
\footnote {UMR 8549: Unit\'e Mixte du Centre National de la Recherche Scientifique
 et de l'\'Ecole Normale Sup\'erieure}\\
24, rue Lhomond, F-75231 Paris Cedex 05, France }
\maketitle\abstracts{
 Recent measurements of the force versus extension curves in stretched single stranded DNA,
under conditions  where the hydrogen bonding  between complementary
bases is inhibited,  provide a  new handle for the study  of
 self-avoiding  effects in  a flexible polymer.  We report in this paper  upon analytic computations
 of  the force versus extension curves  within a continuous version    of  the freely joining chain
  model, with monomer-monomer repulsive interactions.  The problem is formulated
  as a Statistical Field Theory model, endowed with a fixed cutoff  associated with
 the curvature of  the extension versus force curve,  in the free polymer limit. Using a Field Theory
 version of the Hartree-Fock  approximation,  the self-avoiding single polymer
 problem  reduces  to  the solving of a one-dimension self-consistent   integral equation.
 Taking   the short range   potential limit, we obtain    an explicit analytical solution,  involving
the roots of an  algebraic equation. In the low force regime, we find that the slope of the  relative extension
 at zero force increases steadily with the total  monomer  number  $N$, while  it stays
constant for a free polymer.   We  compare, for  $ N=100$,  our analytic  results   with those obtained
by  Monte Carlo simulations   of a  discrete  freely joining chain, in presence
of excluded volume effects. Despite a  difference  between  the repulsive
potential ranges $d$, an appropriate choice  of our single free parameter   leads to a good agreement
between the two computations.  This may be considered as  an indication  that the shape of the extension curve is
not very sensitive to  $ d $, if it is   equal to or less than the monomer length.   }

\section*{Introduction and Summary }
During the last few years, the elastic properties of single biopolymer molecules,
 under physical conditions close to those encountered in living organisms, have been the subject
of extensive  investigations \cite{smith,perk,strick}. In particular,
 the force versus extension curves of double  stranded DNA (dsDNA) have
been measured accurately within  a wide range of pulling forces. In absence of supercoiling constraints, the entropic elasticity
 of the dsDNA  molecule  is very well described \cite{marsig} by an elastic string \cite{fixman} model,
 the  so-called Worm Like Chain (WLC) model, involving a single rigidity 
associated  with  the resistance against the polymer bending.
 The dsDNA persistence length  $ A$ is given, within the WLC model, 
 by the  bending rigidity written with appropriate units.
  Micromanipulations  experiments  lead  to  values clustering around  
  $ A \simeq 50\, nm $, which corresponds to 150  base pairs, hence
  the qualification "stiff" polymer sometimes given to  the dsDNA.
The monomer-monomer  interactions in dsDNA
are  dominated by the screened  Coulomb repulsion and they do not seem to lead to significant corrections to the
WLC model predictions\cite{bouc99}, with  the exception of very long chains, longer than  $50  \,\mu m$\cite{strick00}.

There has been recently a growing interest for  micromanipulation
experiments \cite{bmaier,mndessin} involving the stretching of  single stranded (ss)DNA.
The elasticities  of ssDNA and dsDNA are expected to be different,  for at least two reasons.
One,  despite the lack of   any accurate   determination, it is commonly  believed that
the   ssDNA persistence lies in the range 0.7 to 3 $nm$\cite{kuznet}, corresponding
 roughly   to a base pair number between  2 and  7.
In presence of thermal fluctuations,  the molecular local axis loses the  memory of its
 direction after few base pairs. We express this fact by saying that the ssDNA is a "locally flexible" polymer".
 In such a case,  the WLC model does not lead to an adequate
 description   of the "free"  ssDNA  molecular chain. 
 A more realistic picture is provided by  the freely joining chain (FJC) model,
where the relative direction of two adjacent elementary links 
 is allowed to fluctuate  freely.  As a consequence  of  the ssDNA   high
flexibility,  the self avoiding-effects are  expected to be  more important
 than in  the case of a "stiff" polymer like dsDNA,
 if  the  physical environment is such that   the range of 
the repulsive  monomer-monomer interaction   is about few  $nm$.
The second   reason  for a difference  between the  ssDNA and dsDNA  entropic elasticities  is obviously
the attractive interaction  between complementary  bases, leading   to the so-called "hairpin" structures
 which have  to be opened up  in order to  stretch  the molecular chain. A  theoretical analysis of this
 mechanism  has been performed recently \cite{mezmont}, with the help of the "rainbow" approximation.
The results are in good agreement with the observations of ref.\cite{mndessin} when the buffer
solution contains cations  which  stabilize the hairpin structures.

  Conversely, ssDNA stretching experiments \cite{mndessin} have been  performed
under physical  conditions where   hydrogen bonding  interactions
between complementary  bases  are    suppressed, leaving the  screened  Coulomb  repulsion as
the dominant monomer-monomer interaction. The necessity of accounting properly for self-avoiding
effects  appears  clearly  from  the lack of agreement between  the experimental data  and the
  ideal polymer model   predictions, like the FJC model.
 This  is confirmed  \cite{mndessin}  by a Monte-Carlo  (MC)   simulation
  of the force vs extension curves  within    a  FJC model with excluded volume effects.

  In this paper, we  compute,  by  Statistical Field Theory
  methods,   the    entropic elasticity  of locally flexible polymers  in presence
of repulsive monomer-monomer interactions.

  As a first step (section 1), we have constructed a continuous version of the FJC model,
 taking  the monomer number $n$
  as  the running variable along the chain.   It is well known\cite{bouc99} that one can map the free polymer
  problem upon a single particle quantum problem in an Euclidian space-time. In the present
  case,  the imaginary  time  is just  $-i \times n $. The partition function matrix , relative
  to arbitrary states of the molecule free ends, is   given by $ \hat{Z}=\exp- N \, \hat{ H}_0 $ ,
 where    $ \hat{ H}_0 $  is   the Hamiltonian  of  the attached quantum problem and $N$ the total
 monomer number. In the case
 of the continuous version of FJC model, the Hamiltonian   is  purely kinetic  in the zero
 force limit: $ \hat{ H}_0 =h_0\(( {  \hat{\vec{p}}  }^2  \))  $,  where     $\hat{\vec{p}} =  -i\, \vec{\nabla } $
 is the  momentum operator  conjugated to the vector $ \vec{r}  = {\vec{r}}_N-{\vec{r}}_0$, 
joining the two ends of the chain.
The force stretching energy  $ -\vec{r}\cdot \vec{F}  $ is  then implemented through
the replacement: $   \hat{\vec{p}} \rightarrow   \hat{\vec{p}}-i\vec{F}$.
 One proves easily the following simple relation between the relative extension vs force function $ \zeta_0(F) $
and the  fictitious  free particle Hamiltonian: $  \zeta_0(F)=   2 F  h^{\prime}_0 ( -F^2) $.
  In principle,  the Hamiltonian     $ \hat{ H}_0 $
  is then   obtained by integrating   the extension vs force function   along a finite interval along
 the imaginary $ F$ axis. Unfortunately, the things are not so simple since
    $ \zeta_{FJC}( F )$    has poles along integration path. We have been able
 to bypass this difficulty by building  a  function   $ \zeta_{r}( F )$ regular along the imaginary $ F $ axis
  and differing  from  $ \zeta_{FJC}( F )$    by less  than  a few  $ \% $ along the  whole real $F $ axis.
The coincidence is even better than  $  1 \% $
  in the low   to medium force regime where the self-avoiding effects are the  most important.
 The Hamiltonian     $ \hat{ H}_0 $
 is written finally as   $ \hat{H}_0 ^{r} = h_0 ^{r}\(( (\hat{\vec{p}}-i\vec{F})^2 \))  $  where
 and    $ h_0^{ r}( p^2) $   is   an analytic
 function of $ p $ increasing faster than $p^4$.  In our construction, the quartic expansion
$  h_0 ^{(4)}(p^2) = \frac{b^2 \, p^2 }{ 6}+  \frac{b^2 \, p^4}{ 180} $
 is  in fact  obtained by integrating the cubic expansion of     $ \zeta_{FJC}( F )$, where $b$
 is  the length of  the FJC elementary  link.
  The quadratic  term  corresponds to  the Gaussian model; when it is   used in  association
 with a  short range (Dirac $ \delta \,$function)   momomer-monomer repulsive interaction, it is called  the Edwards model.
   The quartic terms provides
 a  natural cutoff     $\Lambda = 30 b^{-2} $ for the divergences appearing in the perturbation expansion
  of the Edwards model with
  respect   to the monomer-monomer  interactions.  
It will turn out that in the exploration of the low to medium force range
  $ 0 \leq \beta \, F \, b \leq 1.5 $   the quartic Hamiltonian $  h_0 ^{(4)}(p^2) $
 can be substituted to     $ h_0^{ r}( p^2) $, allowing for analytic computations.

  In section 2 we develop an Hartree-Fock treatment of self-avoiding effects  within our continuous version of the FJC
  model. With the help of a well known functional  integral identity\cite{parisimecstat}, the self-avoiding polymer
  problem is   { \it exactly } mapped into that of a quantum single particle moving in a stochastic  imaginary potential,
 described by  the Hamiltonian:  $    \hat{H}^r=  \hat{H}_0 ^{r}  +i \,g\,\phi(\vec{r})   $.
    The functional probability measure  of  the field   $\phi(\vec{r}) $  is of the Gaussian type,  specified  by  the
    two fields product average:
 ${\langle \; \phi (\vec{r_1}) \;\phi (\vec{r_1})\;\rangle }_{\phi} = V( \vert\vec{r_1}-\vec{r_2} \vert)  $
    where  $  g^2 \, V( \vert\vec{r_1}-\vec{r_2} \vert)  $   is   the repulsive monomer-monomer  potential.
     The  partition function   of  the  self-avoiding polymer with fixed free ends is   given by the  stochastic field
     average: $   Z ( {\vec{r}}_N-{\vec{r}}_0,\vec{ F} ,N)=
     \langle \vec{r_N} \;{\vert \langle \exp -N \(( \hat{H}_0 ^{r} 
 +i \,g\,\phi(\vec{r}) \)) \rangle}_{\phi}\,\vert\vec{r_0}\rangle $.
    To proceed, it is convenient to perform the   Laplace transform  of the
      partition function   at zero force  $   Z ( \vec{r},0,N)  \rightarrow  z( \vec{r },\tau ) $,
          where  $ \tau $ is the  variable conjugated to $N$. The result takes the following  simple form,
       $ z( \vec{r },\tau )= 
  \langle \vec{r}  \,\vert {\langle \((  h_0^r\(( {  \hat{\vec{p}}  }^2  \))
 +i \,g\,\phi(\vec{r}) +\tau \))^{-1} \rangle}_{\phi} \, \vert \, 0 \rangle  $,
      which  can be readily written as a Feynman  diagram expansion in power of $g^2$. 
It allows us to use some of the Quantum  Field Theory
      machinery, in particular  the  Dyson integral equations.
   Following a  standard  procedure \cite{fetwal}, we introduce   an approximate  kernel    
in the    Dyson integral equation for the propagator.   We arrive in a rather direct way to the   
  Hartree-Fock equation which performs the exact  summation of  the "rainbow" diagrams.
    The self-consistent  integral equation  takes a remarkable simple form   in the Fourier space.
     It involves then  the Fourier  transform  $\tilde{z} ( \vec{p },\tau ) $ of  $ z( \vec{r },\tau ) $,
   where $ \vec{p}$ is  the  momentum  conjugated  to the relative free end coordinate  $\vec{r }$.
  \be
     \frac{1}{  \tilde{z} ( \vec{p },\tau ) }=   \frac{1}{  \tilde{z_0}  ( \vec{p },\tau )  }+
\frac{g^2}{2 \, \pi^2 } \int d^3 q \, \tilde{z} ( \vec{q },\tau )\,
   \tilde{V}(\vert   \vec{p }-   \vec{q}  \vert )    \nonumber
   \ee
   where $\tilde{V}(\vert \vec{q}  \vert ) $ is the Fourier transform  of the potential $ V(\vert\vec{r} \vert)$
 and $\tilde{z_0} ( \vec{p },\tau )$  corresponds to  the free polymer partition function.

In this exploratory paper we have used a zero range potential, proportional to a Dirac $ \delta $ function. The Hartree-Fock
integral equation reduces in this limit to a numerical equation. We will show that $ \tilde{z} ( \vec{p },\tau ) $
can be written as $( h_0^r( p^2) +\mu(\tau))^{-1} $, where the function $\mu(\tau)$  obeys the self-consistent  equation:
$ \mu(\tau) = \tau + \frac{\lambda}{6}  F( \kappa ,6 \mu(\tau))   $, where  $\kappa =b^2 \,\Lambda^{-2}$ and $\lambda $
a  positive dimensionless constant measuring the strength of the repulsive monomer-monomer potential. In the low to medium
force regime,  $F( \kappa ,\tau)$  is  a rather simple algebraic function of $ \tau $.
The extension versus vs force function  $ \zeta(F,N) $ for the
self avoiding polymer is then  obtained by taking the logarithmic derivative
of the inverse Laplace transform  of $( h_0^r( -F^2) +\mu(\tau))^{-1} $.
The involved integral  is  dominated by the residues of the poles  at the roots of $h_0^r( -F^2) +\mu(\tau)=0$.

   In the low force regime $ 0\leq \beta \, F\,b \leq 0.5  $, 
the computation of the extension vs force function $ \zeta(F,N)$ can
be  performed  analytically up to the very end.  
The final  expression is not   by itself very illuminating,  so  we have
 plotted   a set of extension curves  for a wide range of monomer number: $50 \leq N \leq 5000 $. The most
remarkable feature is  the steady increase with $N$ of the relative  extension vs force slope at the origin ($F=0$).
This contrasts with the case of a free polymer where this quantity is independent of  $ N $. This   behaviour
is also predicted  in the Edwards model  using   scaling arguments\cite{PGDG}, but the effect is notably smaller than in our approach.
Such a difference was to be expected since Renormalization Group  methods cannot be applied to our model which
 is endowed with a fixed cut-off,  associated with the  extension vs force  curvature of the free polymer.

The section 4 is devoted to a comparison of the analytical results of our Hartree-Fock method with those obtained  by a
MC simulation  \cite{mndessin} of a discrete freely joining  chain,  with excluded volume effects. In both cases, the total number $N$
of monomer is one hundred. Our analytic treatment is strictly valid in the low to medium force range  $ 0 \leq \beta\, F\, b \leq 1.5 $,
but we have extended our results to higher forces with the help of  a reasonably safe extrapolation  procedure. We are able in this
way to reach  the force domain, $ \beta\, F\, b \geq 3.5 $, where self-avoiding effects are disappearing. An appropriate
choice of our single adjustable parameter, $\lambda =0.7 $,
 leads to a predicted extension vs force  curve which is in remarkable
agreement with the MC  simulations. It should be stressed  that there is a notable  difference
between the physical inputs  in the two approaches. In our exploratory computation, we use a zero range potential  while
the MC simulations were performed with an excluded volume radius $d$  equal to the monomer length $b$.
 This may suggest that the shape of the extension vs force curve is
 not very sensitive the value  $d$ of the repulsive potential range if $ d\leq b $.
\section{ A continuous formalism for a free flexible polymer. }
In this paper, the  partition function for a stretched  "  locally flexible polymer"  will be  given, in absence
of self-avoiding effects,  by the following path  integral:
\bea
Z_0 ( {\vec{r}}_N, {\vec{r}}_0,\vec{F}) & =
 & \int {  \cal D}[\vec{r}] \exp - \int _0^N \,dn \, {  \cal  E}_0 (n) \;
\nn \\
   {  \cal  E}_0 (n)  & =
  &\epsilon_0 \, \((    (\frac{ d\vec{r} }{ dn} ) ^2 \))-\vec{F} \cdot { d \vec{r} \over
dn}
\label{parfon}
 \eea
where   $ n $ stands for  the monomer number which is treated as a continuous variable
as a result of a coarse-graining of the polymer chain ; ${  \cal  E}_0 (n)$ is  the elastic energy
per polymer written  in thermal energy unit $ k_B T$.The  function
   $\epsilon_0(x)$ is assumed here  to be analytic.
   The equation (\ref{parfon})  could  be  considered    as  the  definition  of what
  we  mean    by a   "  locally flexible polymer" .
 In the   literature about  polymer self-avoiding effects,
the  function $\epsilon_0(x) $ is often taken to be linear:
$  {  \cal  E}_0 (n)=\frac{3}{2 \,b^2}  ({ d \vec{r} \over dn} )^2 -\vec{F} \cdot { d \vec{r} \over dn}$
where $b$ is  the root  mean square distance of two adjacent  polymers.
 The difficulty   with this model- known as the "Gaussian model"-  lies
 in the  fact that  the  extension  vs force curve turns out to be a straight line,  which is quite unrealistic.
The philosophy taken up in  this paper is  to consider  as  the true starting point the
extension vs force function in absence  of  self-avoidance : $  <  z_N > /N  = \zeta_0(F) $,  rather than
 the elastic energy $ \epsilon_0 \, \(( ({ d \vec{r} \over dn} )^2 \)) $
 appearing  in equation (\ref{parfon}).
\subsection{ An Hamiltonian formalism to relate the extension vs force  function  to  the linear elasticity density.}
We are now going to write the discretized  form of the partition function  given in equation
(\ref{parfon})  in terms of   the transfer matrix operator $  \hat{T_0} $ :
\bea
Z_0 ( {\vec{r}}_N, {\vec{r}}_0) & =
 &  \langle \vec{r}_N  \vert \, {\hat{T_0}}^N \,\vert  \vec{r}_0 \rangle,   \nn \\
\langle \vec{r}_{n+1} \vert \, \hat{T_0} \,\vert  \vec{r}_n \rangle &=
 &  \exp-\((  {\epsilon}_0 \(( (  \vec{r}_{n+1} - \vec{r}_{n} ) ^2 \))-\vec{F}\cdot (\vec{r}_{n+1}
-\vec{r}_{n}) \)).
\label{tmr}
\eea
  The operator $ \hat{T_0} $   is invariant upon space translations  and  is then expected
   to   be diagonal   within   the plane wave basis:
$ \langle \vec{r} \,\vert \,\vec{p} \rangle = (2\pi)^{-\frac{3}{2}}\exp(i \,\vec{p}\cdot \vec{r}) $. The transfer matrix
in the momentum space is  then readily obtained :
 \be
 \langle \vec{p}_{n+1} \vert \, \hat{T_0} \,\vert  \vec{p}_n \rangle =
 {\delta}^3( \vec{p}_{n+1}-\vec{p}_{n}) \, {\tau}_0\(( (\vec{p}_{n} -i\vec{F} )^2\)),
\label{ trmpa}
\ee
where $ {\tau}_0\(( \vec{p}^2  \)) $    is  given in terms  of $\epsilon_0 \((\vec{r}^2  \))$
    by the  following Fourier integral:
\be
{\tau}_0\(( \vec{p}^2  \)) =  \int\, d^3 \, r \exp i \,\vec{p}\cdot \vec{r}
 \;\exp\((-\epsilon_0 ( \vec{r}^2) \)).
\label{trmpb}
\ee
 Introducing the momentum operator $ \hat{\vec{p}}=-i\, \vec{\nabla } $ the transfer
matrix operator  can be written  under the simple form:
 $ \hat{T_0} =     {\tau}_0\((  (\hat{\vec{p}}-i\vec{F})^2\)) $.
 We arrive in this way to a compact
    formula for the partition function,  which will  lead us  to the Hamiltonian
$ \hat{ H}_0$  of an associated quantum mechanical  problem:
\bea
 Z_0 ( {\vec{r}}_N- {\vec{r}}_0,\vec{F},N) &=
  &  \langle \vec{r}_N \vert \,   {\tau}_0^N\((   (\hat{\vec{p}}-i\vec{F})^2 \)) \,\vert  \vec{r}_0 \rangle =
\langle \vec{r}_N \vert \,\exp(-N \hat{H}_0)\,\vert  \vec{r}_0
\rangle,  \nn\\
\hat{H}_0 &=&  h_0\(( (\hat{\vec{p}}-i\vec{F})^2 \)) =
-\log\, {\tau}_0\(( (\hat{\vec{p}}-i\vec{F})^2  \)),
\label{effhamil}
 \eea
where in  our writing  of the partition  function we have made explicit the fact
 that  $ Z_0$  is invariant under  space translation. In the simple case of the
Gaussian model,  one finds: $ \hat{H}_0 =- \frac{b^2}{2} \,(\vec{\nabla }+ \vec{F})^2 $.

Let us  proceed   to the evaluation of  the  relative extension vs force function $ \zeta_0(F)$.
The extension average $ < \vec{r} >$ is given by :
\be
 < \vec{r} >=\frac{ \partial  }{ \partial \vec{F} } \log\(( \int d\, r^3 Z_0 ( \vec{r} ,\vec{F},N) \))=
 -N   \frac{ \partial  }{ \partial \vec{F} } \,h_0( -\vec{F}^2).
\ee
Taking the stretching force along the $ z $ axis, one gets the relation which  will allow us  to construct
 the  effective Hamiltonian $\hat{H}_0 $  from  the  extension  vs force curve $ \zeta_0(f) $:
\be
\zeta_0(F)=  \frac{< z >} {N}= 2 F  h^{\prime}_0 ( -F^2)=- 2 f  \frac { {\tau}^{\prime}_0 ( -F^2) } {\tau_0 ( -F^2) }.
\label{extvsf}
\ee
Let us now consider the case of the freely  joining  chain (FJC) which could be considered as a
reasonable starting point  for the study of  self-avoiding effects    for a  " locally flexible" polymer.
The transfer matrix  operator   for  a chain of  freely  joining  monomers,  having a  fixed length $b$,
can be written as:
$ \langle \vec{r}_{n+1} \vert \, \hat{T_0}(FJC) \,\vert  \vec{r}_n \rangle=
\delta\(( \vert\vec{r}_{n+1}-\vec{r}_{n}\vert -b  \))  \,\exp \,\vec{F}\cdot \vec{r} $.
 It can be easily written in momentum space  form  as in eq. (\ref{ trmpa}) by taking:
${\tau}_0\(( \vec{p}^2  \)) = \sin(p\, b)/( p \,b )$.   Inserting this result  in eq. (\ref{extvsf}), one gets
immediately  the  FJC model  extension  vs force  curve in terms of the  reduced force $ f= Fb $ :
\be
 \zeta_{FJC}(f)= b \((\coth( f) -\frac{ 1}{ f}\)) = \frac{ f}{ 3 }-  \frac{ f^3 }{ 45 } +O( f^5).
\label{extFJC}
\ee
\begin{figure}
\centerline{\epsfxsize=120mm\epsfbox{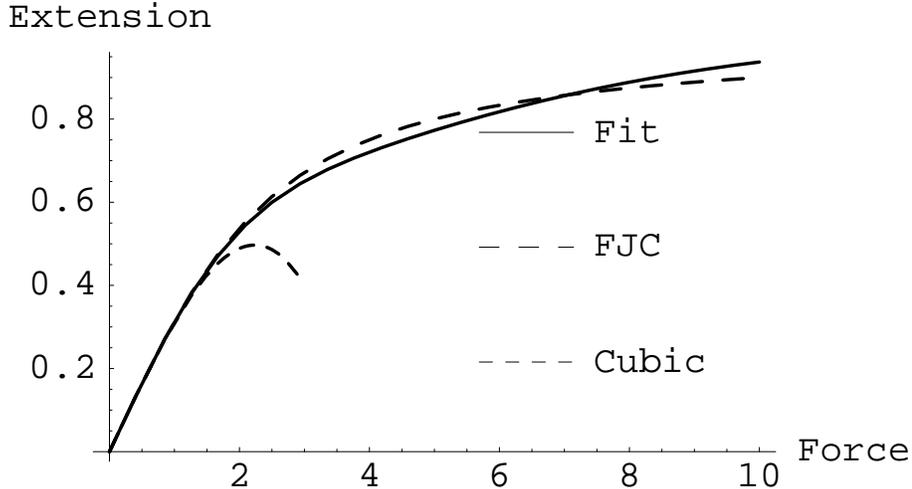}}
\caption{ \footnotesize  This figure  gives the results of  a fit  of the extension vs force function  $\zeta_{FJC}(f)$
by a superposition of two error functions. The idea behind this procedure is to build
a continuous variant   of the FJC mode,  where the associated free Hamiltonian  is regularized, while
keeping the essential features of the FJC model elasticity.
We see  clearly  on this figure that  this goal has been successfully achieved.
 We have required the fitted function  $\zeta_r(f)$, with the label " Fit"to coincide with $\zeta_{FJC}(f)$  to third order
in the reduced force $ f=F/b $. The dotted curve, with the  label   "Cubic", represents the third order
polynomial expansion  of $\zeta_{FJC}(f)$.
 }
\label{fig1}
\end{figure}
\subsection{ Construction of  an  effective  regular Hamiltonian  from  the extension  vs force function. }
If one applies blindly the formula (\ref{effhamil}) in order to get the effective Hamiltonian
$\hat{H}_0 $, one gets $ h_0 (p^2)= -\log(\frac{\sin p\,b } {p\,b} ) $, which is clearly singular when
$ p\,b = n \,\pi $, $n$ being an arbitrary non zero integer. One may  try  to regularize  the model by replacing
in  $ \langle \vec{r}_{n+1} \vert \, \hat{T_0}(FJC) \,\vert  \vec{r}_n \rangle $ the $\delta $
function by a Gaussian, allowing for fluctuations of the monomer length  $b$. This does not solve really the
problem since $ {\tau}_0 (p^2) $ still oscillates around zero and $ h_0(p^2)$ does not increase
fast enough  with  $p^2$. The fact that the above formal derivation  of $ h_0(p^2)$    leads to a singular result  should not be
considered as a surprise. A typical  raw  FJC contour line,  in absence of any  coarse-graining, is very far
from   what  could be obtained by  the  discretization   of  a 3D-curve  with  smoothly varying
 derivatives  of  the running coordinate $ \frac{d\vec{r} }{n} $.

We have adopted here  the view that
there is  nothing sacred about the FJC  model: it is just a guide to get a reasonable extension vs force curve
for a "locally flexible" polymer.

 Let us sketch the construction of  a regular Hamiltonian  $ h_0^r (p^2)$ associated with  an extension vs force
 function $ \zeta_r(f) $     differing from $\zeta_{FJC}(f)$  by  less than few $ \% $  for the whole range of values  of the
  reduced force $f$. The result is exhibited on Fig.\ref{fig1}. A convenient   starting point
    is the  derivative  $\zeta^{\prime}_{FJC}(f)= -\sinh^{-2}(f)+f^{-2} $. It is a  bell-shaped  even function of $f$ we  write
    as $ \chi(f^2)$. In the case of the FJC model    $ \chi(z)$ is an analytic function of $ z $
    with an infinite set of poles at $ z_n= - ( n \pi )^2 $,
    which are responsible for the  singularities of $ h_0(p^2) $. We shall  suppress them by replacing     $\chi_{FJC}(z)  $  by
  a     function  regular along the whole $ z $,   having the general form:
$\chi_{r}(z)  = \int_0 ^M d\, m \, g(m) \, \exp(-m \, z)$. Here $ g(m) $ is a distribution  which is non zero  if and only if
  the  real  number  $m$   belongs to the finite domain  $  0 \leq m \leq M $ where $ M $ is {\it a finite positive real}.
Note that this   representation is not valid for      $\chi_{FJC}(z)  $ unless $M$  is infinite.

  As in all regularization methods, there is always some arbitrariness. Two criterion's have  to be used as guiding principles:
  one is  mathematical simplicity,  the second is  keeping  to a minimum the number  of { \it ad hoc } parameters.
  We have found that  a very simple form  of the distribution  $ g(m)$ is adequate for our purpose:    $\chi_{r}(z) $
 is taken as the sum of two exponentials.
   Returning to the $f$ variable,  it  implies that   the  regularized  bell-shaped  curve $\zeta^{\prime}(f) $
 is the sum of two Gaussian.  The main effect of the
  regularization is to change the shape of the high force tail of  $\zeta^{\prime}(f) $  in a domain where self-avoiding
effects  are becoming very weak.  One of the peculiar
  feature of the  analytic continuation $ f \rightarrow i \, p $  is that a modification performed  for large  positive $f$  values  can
   induce or suppress singularities
  for finite values of $ p $ : $ \pi , 2 \pi $ .....

  By performing a   simple quadrature, we arrive  finally at the following expression
 for the  regularized extension function $  \zeta_r(f)$ :
 \be
 \zeta_r(f)=  \frac{b}{2}\(( ( 1-c)\,{\rm erf}(a_1\,f)+ ( 1+c)\,{\rm{erf}}(a_2\,f) \)),
\ee
where  $  {\rm erf} (z)= \frac{2}{\sqrt{\pi}} \int_ 0^{\infty}  d\,t \, \exp-t^2 $.
Note that if  $a_1$ and  $a_2$ are positive, then $ \zeta_r(\infty)=1 $.
The determination  of the parameters $a_1$, $a_2$ and $c  $   is performed in two steps:

\textbf{ Step 1. } $a_1$ and  $a_2$  are obtained  as functions of $c$
by requiring the  equality of the third order expansions of $ \zeta_r(f) $ and  $\zeta_{FJC}(f)$ with respect to $f $.
As it is clear from    Fig.\ref{fig1}, this  constraint    insures
 that    $ \zeta_r(f) $ and  $\zeta_{FJC}(f)$  coincide in the  low to medium force
range   $ 0 \leq\,f  \, \leq 1.5 $, where self- avoiding effects are expected to  be important.

\textbf{Step 2.}  The parameter   $ c $   results   from a minimization of  the mean square average  difference  :
 $ \Delta_2(f_m)= \frac{1}{f_m}  \int_0^{fm} df \,  (\zeta_r(f)- \zeta_{FJC}(f))^2 $. For $ \vert f  \vert \geq 10 $
the molecule extension   is above   $ 90 \% $  of its   maximum.
Self-avoiding  effects  are  thus  expected to be small, and this  is confirmed by a look at the curves displayed on FIG. \ref{fig6}.
 This justifies our choice:  $ f_m=10 $.
 The  resulting  numerical  values  for  the constrained parameters are then given by  :
\be
    c= -0.035 \; ,\; a_1= 0.093 \;,\; a_2= 0.484
\ee
With the above numbers, one gets for the root mean square difference :  $ \sqrt{ \Delta_2(f_m)} = 0.0198 $.
 Such a small difference  gives  a measure of the  success   of our enterprise to suppress   the singularities of $ h_0(p^2) $   while
  keeping to a few percent level   the modifications of the FJC model  extension vs force function. It
  is also important to note that there is {\it  no free parameter left. }

\begin{figure}
\centerline{ \epsfxsize=80 mm \epsfbox{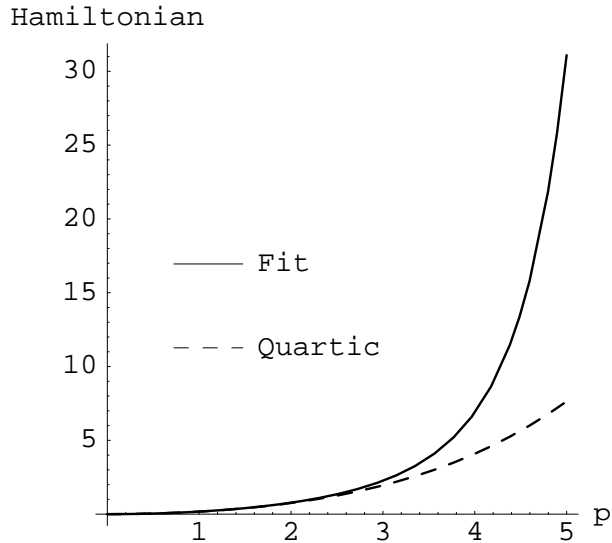} }
\caption{\footnotesize  On this  figure, the solid line   represents  the effective Hamiltonian   $ h_0 ^{r}(p^2)$,
 obtained by a simple
quadrature  from  the  analytic continuation  $ f \rightarrow i \,p  $ of the  force -extension curve $\zeta_r( f)$. The
dashed curve  gives the fourth order expansion:
 $ h_0 ^{(4)}(p^2) = \frac{b^2 \, p^2 }{ 6}+  \frac{b^2 \, p^4 }{180} $.  It nearly
 coincides  with  $ h_0 ^{r}(p^2)$ in the interval $ 0 \leq  p\leq 3 $  and will be used
  to get the analytic results given in this paper.
   Note  the  fast increase of two the curves for large values
   of  $p^2$. It  guarantees the absence of divergences in the field theory description of self-avoiding effects which
   will be  adopted  in the present work.
   }
  \label{fig2}
  \end{figure}

We are now ready to write down the effective  Hamiltonian $ \hat{H}_0 ^{r}$ which will  be used in this work :
\bea
\hat{H}_0 ^{r}&=& h_0 ^{r}\(( (\hat{\vec{p}}-i\vec{F})^2 \)),  \nonumber \\
 h_0 ^{r}(p^2) &=&-\int_0^{p \,b} du \, \zeta_r(i\,u)\nonumber \\
 & =& \frac{1}{2}(1-c)\(( \frac{a_1}{\sqrt{\pi}} (1- \exp(a_1\, p^2\, b^2)+p\,b \, {\rm  erfi}( a_1 \,p\,b)\))      \nonumber \\
   & & +     (c \to -c \, ,\, a_1 \to a_2 ),
\label{hfit}
\eea
where $ {\rm  erfi}(z)=  {\rm erf}(i \, z)/i $. The function $ h_0 ^{r}(p^2)$  is displayed on   Fig.\ref{fig2}.

We have  verified that the use of $ h_0 ^{r}(p^2)$ does not introduce  any significant
nonphysical artefact  in the model.
To do that, we have worked backward:
taking $ h_0 ^{r}(p^2)$ as starting point,   we have  computed  the probability distribution of
 $ l_n=\vert \vec{r}_{n+1} - \vec{r}_{n} \vert $, given by
$ {\cal P} (l_n)= l_n^2 \, \tilde \tau( l_n) ={l_n}^2\, \exp - {\epsilon}_0 (l_n^2)$.
Using eq.(\ref{trmpb}), we get  $\tilde \tau( l_n)$  by   performing the $3D$
inverse Fourier transform of  $ \exp- h_0 ^{r}(p^2) $. Discarding   a  very weak damped  oscillating tail at  large $l_n$,
{\it having a maximum amplitude which stays below the $ 1.5 \% $ level}, we get a well behaved  Gaussian-like
bell- shaped  curve centered at $ l_n \approx b$ with a half-width $ \Delta \,l_n\approx b/2 $.
In our procedure,  regularity conditions are enforced at each step. Our derived  probability
distribution   $ {\cal P} (l_n) $  is not  expected  to be  a Dirac $\delta $ function  but should
   rather  be  a peaked   curve with a finite width,
   which should be  of the order of   the coarse-graining length  resolution needed to wash  out the strong fluctuations  of
the raw FJC  contour line.

  The quartic expansion of $ h_0 ^{r}(p^2)$  is, by construction, independent of the parameters
$ a_1,a_2 $ and $c$ :
\be
h_0 ^{(4)}(p^2) = \frac{b^2 \, p^2 }{ 6}+  \frac{b^2 \, p^4 }{ 180}
\label{hquart}
\ee
As it is apparent on Fig \ref {fig1} and \ref{fig2}, it could constitute an adequate starting point  for the study  of the
self-avoiding effects of the FJC model  in the force range $ 0\le f \le 1.5 $,  which will  be studied in the next sections.
The effective Hamiltonian $\hat{H}_0^ {(4)} $ associated with  $  h_0 ^{(4)}(p^2 ) $ can be viewed as a cutoff
version of the Gaussian model. Indeed,  the corresponding Green function in  the momentum
space: $\tilde{G}_0 (p^2,z)= (z -  h_0 ^{(4)}(p^2) )^{-1} $ behaves as $ p^{-4} $  when
$ p^2  \gg \Lambda^2 = 30 \, b^{-2} $. The cut-off $\Lambda$ is not here an {\it  ad hoc}  ingredient,
since it is related to a  simple physical quantity: the curvature of the extension vs force curve
 in the low force regime.

 We would like to stress that      $ h_0 ^{(4)}(-f^2) $ is not valid beyond $ f \geq  2$.  Higher polynomial
 expansion of  $ h_0(p^2)=-\log(\sin(p)/p)$,  as a substitute for  $ h_0^r(p^2) $, will lead to
 difficulties due to  the divergence of the  entire series expansion  of     $ h_0(p^2)$ at $ p=  \pi, 2 \pi ... $

\section{ The Hartree-Fock approximation  for Self-avoiding   Polymers.}
Let us incorporate in  the partition function the  self-avoiding effects   induced by  the  monomer-monomer repulsive central potential
  $ g^2\, V( \vert \vec{r}_1 - \vec{r}_2  \vert)$ :
\be
Z ( {\vec{r}}_N-{\vec{r}}_0,\vec{F},N)  =
  \int {  \cal D}[\vec{r}] \exp - \int _0^N \,dn \, \((\,{  \cal  E}_0 (n) +\int _0^n \,dn^{\prime}\,
g^2 \,V( \vert\vec{r}_n - \vec{r}_{ n^{\prime}}\vert ) \)).
\label{partfselfavoid}
 \ee

Before entering into the details of our approximation scheme, it useful to derive a  convenient  formula
for the evaluation  of  the chain  extension vs force. We  introduce the space average   partition  function $\bar{Z}( \vec{F}^2,N) $
 obtained by integrating upon the distance $\vec{r}={\vec{r}}_N-{\vec{r}}_0$
between the two free ends of the molecular chain.:
\be
\bar{Z}( \vec{F}^2,N)=\int d \,r^3 \,Z (\vec{r},\vec{F},N)= \int d \,r^3 \, \exp( \vec{r} \cdot \vec{F}  ) Z (\vec{r},0,N),
\label{zedbar}
\ee
where have factorized  the stretching  potential energy and  exploited
 the rotation invariance  of $Z (\vec{r},0,N)$. Taking the stretching force along the $z$ axis, $ \vec{F}= \hat{z} \, F$, we can
readily obtained  the relative extension vs force expression :
\be
  \zeta(F,N)=  \frac{ < \, z \,>}{N}=\frac{\partial } { N \partial \, F} \log\,\((\bar{Z}( F^2,N)\)).
\ee
The computation  method developed  in this section  will give  in a rather direct way
 the Fourier transform  of the  partition function  of the unstretched polymer:
\be
\widetilde{ Z} ( k^2,N)= \int d\, k^2 \exp(i \,\vec{r} \cdot\vec{k }) \,Z (\vec{r},0,N) .
\label{Zfour}
\ee
It is then of interest to write the extention  vs force formula in terms of this quantity.
To do that, let us take   again $ \vec{F} $ along the $ z$ axis and  perform the analytic continuation $ F \rightarrow i\,p  $
upon the r.h.s. of eq. (\ref{zedbar}).  Inserting  the Fourier expansion of $Z (\vec{r},0,N)$,  one can readily
 integrate over $ \vec{r} $  and arrive at the relation $  \widetilde{ Z} ( p^2, N) =\bar{Z}( -p^2,N) $.
 Performing
the analytic continuation backwards,  one gets the searched-for   formula:
\be
\zeta(F,N)=  \frac{ < \, z \,>}{N}=\frac{\partial } { N \partial \, F} \log\,\((\widetilde{Z}( -F^2,N)\)).
\label{zetasav}
\ee
\subsection{ A Statistical  Field Theory model  for self-avoiding  effects in   flexible polymers .}
Following a standard procedure \cite{parisimecstat,polymvilgis}, we are going to
 transform  the evaluation  of the r.h.s. of (\ref{partfselfavoid} ) into a field theory
problem involving an auxiliary scalar  field $\phi(\vec{r}) $. The  basic  tool is the well known   Gaussian
functional   integral  identity :
\bea
 { \cal E }  [V] &=& \exp- \frac{1}{2} \int _0^N dn \int _0^N \,dn^{\prime}\,
 g^2 \,V( \vert\vec{r}_n - \vec{r}_{ n^{\prime}}\vert)         \\
   &=& (\det\, V)^{1/2} \int  { \cal D}[\phi]
 \exp- \((\int_0^N d\,n \, i \, g \,\phi(\vec{r}_n ) \))   \nonumber \\
& & \times  \exp- \((\int d^3 \,r_1 \, d^3 \, r_2  \, \frac{1}{2}
V^{-1}(\vec{r}_1 - \vec{r}_2 )   \phi(\vec{r}_1) \,\phi(\vec{r}_2 )  \)).
\eea
In   the case of the  Yukawa potential  $ V(r)=\frac{1}{ 4 \pi \,r }\exp-r/a $,  the inverse operator is given
by :
$$ V^{-1}(\vec{r}_1 -\vec{r}_2 ) =\delta^3 ( \vec{r}_1 -\vec{r}_2 ) (-\vec{\nabla }_1 ^2 +a^{-2} ). $$
We arrive in this way to    the Field Theory  formula   for the  partition function upon  which  we are going to
build our approximation scheme:
\bea
Z ( {\vec{r}}_N - {\vec{r}}_0,\vec{F},N)  & = & \int { \cal D}[\phi]
 \exp - \((  \int d^3 \,r_1 \, d^3 \, r_2  \, \frac{1}{2}
 V^{-1} (\vec{r}_1 - \vec{r}_2 )   \phi(\vec{r}_1) \,\phi(\vec{r}_2 )  \))       \nonumber \\        \nopagebreak
  & & \times   \langle \vec{r}_N \vert \,\exp -N \hat{H}( \vec{F},\phi) \,\vert  \vec{r}_0\rangle.
\label{partfieldth}
\eea
 The Hamiltonian $ \hat{H}( \vec{F},\phi) $ involves the stochastic  imaginary potential
$ i \,g \,\phi(\vec{r}) $:,
\be
 \hat{H}( \vec{F},\phi)  =   h_0 \((  (\hat{\vec{p}}-i\vec{F})^2 \)) +i \,g\,\phi(\vec{r}),
\ee
 where  the  "kinetic" term  $ h_0(p^2) $ is to be identified with $ h_0^{(4)}(p^2)$ ( eq. (\ref{hquart})), or eventually with
$ h_0 ^{r} (p^2) $  ( eq. (\ref{hfit})).
\subsection{ A simple Feynman graph expansion  for the Laplace transform of the partition function.}
We  proceed by   performing  the Laplace transform of  (\ref{partfieldth})  with respect to $N $:
\bea
  z( {\vec{r}}_N - {\vec{r}}_0,\vec{F},\tau) &=& \int _0^{\infty} d\,N \, \exp(- \tau \, N)\,
 Z ( {\vec{r}}_N-{\vec{r}}_0,\vec{F},N)
   \nonumber \\
 &= &
     \langle \vec{r}_N \vert \,\langle\((  h_0 \((  (\hat{\vec{p}}-i\vec{F})^2 \)) +i \,g\,\phi(\vec{r}) +\tau
\))^{-1} \rangle_{\phi}\,\vert  \vec{r}_0\rangle ,
\eea
  where we have  introduced  a  compact notation to describe the  average
of  a given functional $ {\cal F} [ \phi] $   over  the stochastic field $\phi(\vec{r})$:
\be
{\langle \; {\cal F} [ \phi] \;\rangle }_{\phi}= \int { \cal D}[\phi] {\cal F} [ \phi] \exp - \((  \int d^3 \,r_1 \, d^3 \,
r_2  \,
\frac{1}{2}   V^{-1}(\vec{r}_1 - \vec{r}_2 )   \phi(\vec{r}_1) \,\phi(\vec{r}_2 )  \)).
 \ee
One sees immediately that the  average product of an odd  number of fields  $ \phi (\vec{r_i})$ is zero.
 We give  some simple examples of even  products:
\bea
{\langle \; \phi (\vec{r_1}) \;\phi (\vec{r_1})\;\rangle }_0 &=& V( \vert\vec{r_1}-\vec{r_2} \vert) \nonumber \\
 {\langle \; \phi (\vec{r_1}) \;\phi (\vec{r_2})\;  \phi (\vec{r_3}) \;\phi (\vec{r_4})\;\rangle }_0 &=&
V(  \vert\vec{r_1}-\vec{r_2} \vert) \,V( \vert \vec{r_2}-\vec{r_4} \vert)+V( \vert \vec{r_1}-\vec{r_4} \vert) \,
V( \vert\vec{r_2}-\vec{r_3} \vert)    \nonumber \\
&  &+ V(  \vert\vec{r_1}-\vec{r_3} \vert) \,V(  \vert\vec{r_2}-\vec{r_4} \vert)
\label{fiprodav}
\eea
The rule is easily generalized to a product of  an arbitrary  even number   of fields and
 leads  readily to a Feynman  graph expansion of $ z(  {\vec{r}},\vec{F},\tau)  $ in power of $ g^2$.
The perturbation expansion  of the resolvant  $ R(\tau)= ({\hat{H}}_0+\hat{U}+\tau )^{-1} $
( the operator  $\hat{U}$ is defined by   $ \langle \vec{r_2}\vert \hat{U}\vert \vec{r_1}\rangle=
i \,g \, \delta^3 ( \vec{r_1}-\vec{r_2}) \, \phi(\vec{r_1}) $ ) is given by the geometrical series,
keeping only the relevant  even terms :
\bea  R(\tau) &=& R_0(\tau) +R_0(\tau) \; \hat{U}\;R_0(\tau) \; \hat{U} \;R_0(\tau)     \nonumber \\
& & +R_0(\tau) \hat{U}\;R_0(\tau) \; \hat{U}\;R_0(\tau) \; \hat{U}\;R_0(\tau) \; \hat{U}\; R_0(\tau) + ... ,
\eea
 where $ R_0(\tau) = ({\hat{H}}_0+\tau )^{-1} $.
 To make   more apparent the analogy  with the Field Theory formalism  we define  the Green functions:
\bea
G_0( \vec{r}_2 - \vec{r}_1 ,\tau) &=&  z_0( {\vec{r}}_2 -{\vec{r}}_1,\vec{F},\tau)=
\langle \vec{r_2}
\;\vert R_0(\tau)\; \vert \vec{r_1}\rangle  \nonumber     \\
  G( \vec{r}_2 -\vec{r}_1 ,\tau) &= &z( {\vec{r}}_2 -{\vec{r}}_1,\vec{F},\tau)=
\langle \vec{r_2} \;{\vert \langle R(\tau) \rangle}_{\phi}\;\vert \vec{r_1}\rangle
\label{greenfon}
\eea
\subsection{ An Hartree-Fock approximation from the Dyson equation. }
Following Dyson \cite{eqdys}, we introduce the proper self-energy  $  \Sigma ( \vec{r}_2 - \vec{r}_1)    $
  obtained  by summing  all self-energy
 diagrams which cannot  be divided  into two disconnected pieces  by cutting the "particle  line"
associated with the Green functions $ G_0( \vec{r}_2 - \vec{r}_1 ,\tau) $.  Pure topological arguments
lead to  the well known  Dyson equation for the propagator:
 \bea
G( \vec{r}_2 - \vec{r}_1 ,\tau) &=& G_0( \vec{r}_2 - \vec{r}_1 ,\tau) +  \nn \\
 & &  \int  d^3r_3   d^3r_4G_0( \vec{r}_2 - \vec{r}_3 ,\tau)
\Sigma ( \vec{r}_3 - \vec{r}_4)  G( \vec{r}_4 - \vec{r}_1,\tau).
\label{dyseq}
\eea

Taken in  their  exact form, the Dyson  equations have  played  a very important role  in the Renormalization
 Theory but they are of little use for practical computations. They are,  however,  very
convenient to devise  non trivial  infinite summation  scheme by taking   an approximate form
of the  proper self-energy.  The  best  known  example of this procedure is
the Hartree-Fock  (HF) method  \footnotemark[2]\footnotetext[2]{ The formalism
used  here  takes its inspiration    from the  Fetter and Walecka  remarkable  book
dealing, with  Quantum Theory of Many Particle Systems \cite{fetwal}.  } which relies upon the approximate formula:
\be
\Sigma ( \vec{r}_2 - \vec{r}_1) \approx  \Sigma^{(1)} ( \vec{r}_2 - \vec{r}_1)=
  -g^2 \, G( \vec{r}_2 - \vec{r}_1 ,\tau)\,
V (\vec{r}_2 - \vec{r}_1)
\label{sigHF}
\ee
A physical interpretation of the HF  approximation can be given by  noting  that
$  {\Sigma}^{(1)} ( \vec{r}_2 - \vec{r}_1)$
 is obtained by replacing,  in the lowest    order  proper  self-energy, the free polymer  propagator
    $G_0( \vec{r},\tau) $  by  $ G( \vec{r},\tau ) $ which incorporates in a self consistent way
higher order monomer-monomer  interactions . Although technically  rather different, the   present HF
method bears a  close  analogy to  the "rainbow"  diagrams summation   method used  to
compute "hairpin" effects in a  single stranded  discretized  DNA chain\cite{mezmont}.
\begin{figure}
\centerline{\epsfxsize=120mm \epsfbox{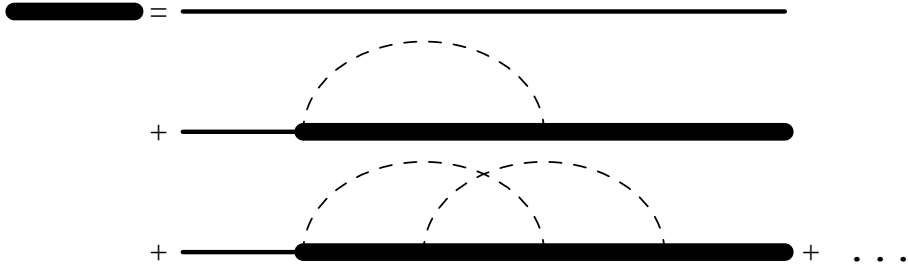}}
\caption{\footnotesize  Graphical   representation  of the Hartree-Fock approximation.
The thin line and the thick   lines  stand respectively for the free  and the self-avoiding
 polymer Green functions.   The dashed half-ellipses represent the self-avoiding  interactions.
 The two upper diagrams schematize the HF approximation
used in this paper which performs the summation  of the planar graphs. An infinite set
of non-planar diagrams  is generated by including the contribution of the bottom diagram.
 }
\label{HFg}
\end{figure}

One of the virtues of the Field Theory  formulation   of the Hartree-Fock  method is to allow a  rather
 systematic treatment of the  correcting  terms. The first step consists in adding to the HF proper self-energy
given by eq.(\ref{sigHF}) a correction term  $ \Sigma^{(2)} ( \vec{r}_2 - \vec{r}_1) $  which will generate
an infinite set of non-planar diagrams. It is constructed from the lowest order { \it non-planar } diagram contribution
to the proper self-energy  in the same way as $ \Sigma_{HF} ( \vec{r}_2 - \vec{r}_1) $
 was obtained   from the  lowest  order proper self energy:
\bea
\Sigma^{(2)} ( \vec{r}_2 - \vec{r}_1)  &= & g^4\int d^3r_3   d^3r_4
G( \vec{r}_2 - \vec{r}_4)V( \vec{r}_2 - \vec{r}_3) \times     \nn   \\
 & & G( \vec{r}_4 - \vec{r}_3) V( \vec{r}_4 - \vec{r}_1)G( \vec{r}_3 -\vec{r}_1).
\eea
We give in FIG.\ref{HFg} a graphical representation  of the H.F  procedure. The thin lines
 stand for free polymer  Green function  $G_0( \vec{r},\tau) $,
  while  the thick ones  correspond to the
self-avoiding  polymer  Green function  $G( \vec{r},\tau) $.  The  calculations performed
in the present   paper incorporate  the  contribution of   the two upper diagrams.  The correction
associated with $ \Sigma^{(2)} ( \vec{r}_2 - \vec{r}_1) $ will be very briefly discussed in the next section.
 \subsection{ The HF equation in momentum space.}
 Because of  the space translation invariance of  eq. (\ref{dyseq})
considerable simplification   is achieved   by working  within  the plane  wave basis where
the  operators $\hat{G}, \hat{G}_0 $ and $ \hat{\Sigma} $  are diagonal   :
\bea
 \langle \, {\vec{p}}_2 \, \vert \hat{G} \vert  \,{\vec{p}}_1 \, \rangle &=  &
 \langle \, {\vec{p}}_2 \;{\vert \langle R(\tau) \rangle}_{\phi}\; \vert
\,{\vec{p}}_1 \, \rangle
 =  \delta^3( {\vec{p}}_1 -{\vec{p}}_2)\,  \tilde{G}({\vec{p}}_1 ,\tau)    \nn \\
 & = &  \delta^3( {\vec{p}}_1 -{\vec{p}}_2)\,
\int d^3\, r \exp(i\,{\vec{p}}_1 \cdot \vec{r} )\, G(\vec{r}).
\eea
The Hartree-Fock integral   equation can then  be written under the rather compact  form:
\bea
  \widetilde{G}({\vec{p}}_1 ,\tau) ^{-1} & =  & \widetilde{G}_0 ({\vec{p}}_1 ,\tau) ^{-1}-
\widetilde{\Sigma}({\vec{p}}_1 ,\tau )  \nonumber \\
& =& \tilde{G_0} ({\vec{p}}_1 ,\tau) ^{-1}
  +\frac{g^2 } { (2\pi^3 ) }\int d^3 p \,\widetilde{G}(\vec{p} ,\tau) \,\tilde{V}( \vert{\vec{p}}{_1}-\vec{p} \vert).
\label{hfinteq}
\eea
   In order to compute  the extension  vs  force curve  in the general case, one has to  proceed as follows:

i) First, we take the limit $ \vec {F}=0 $. The   operators $  R_0(\tau) $  and $ {\langle R(\tau) \rangle}_{\phi}$
are then rotation invariant and $ \widetilde{G}(\vec{p},\tau)= \widetilde{g}(p^2,\tau ) $. The  HF integral equation becomes
one-dimensional  with as kernel $ K(p_1,p)=2\, \pi \, p^2 \int _{-1}^1 d\,x \widetilde{V}\((\sqrt{ p_1^2 -p^2 -2 \, x\,p_1\,p }\))$.
 In the particular case of Yukawa or exponential potentials, $ K(p_1,p) $ is easily computed analytically.
Returning to Statistical  Mechanics notations one sees on eq.(\ref{greenfon})  that $\widetilde{g}(p^2,\tau ) $ is nothing
but  the Laplace  transform  $\widetilde{z}(p^2,\tau ) $ of $ \widetilde{Z}(p^2, N)$, introduced   in eq. (\ref{Zfour}).

 ii) The second  step involves the  analytic continuation  $ p^2 \rightarrow -F^2 $  of $ \widetilde{z}(p^2,\tau )$.
 Then, one has to  perform  the   inverse Laplace transform in order
 to get the extension vs  force curve $ \zeta(F,N)$ from eq.(\ref{zetasav}).
 As we shall see, for the particular  case  of  the $\delta $  function potential, $ \zeta(F,N) $ is dominated
 in the large $N$ limit  by   the  contribution  of   the pole of  $ \widetilde{g}(- F^2, \tau)= \widetilde{z}(-F^2,\tau )$,
 occurring at  the largest algebraic value  of $ \tau $.
\section{ The  HF Method    for a Short Range
Repulsive Monomer-Monomer Potential.}
From now on, we are going to use   $b$  as  the unit of length and as before $ k_B \,T $ as the unit of
energy. The short range limit of the Yukawa potential $ g^2 \,V(r) =\frac{g^2 \,a }{4 \pi \,r} \exp-\frac{r}{a}  $,
is obtained by taking $ a \rightarrow  0 $, keeping   $ g^2 \, a^3 $  finite. The result   is just the $\delta$-function potential :
$ V(\vec{r})=  g^2 \, a^3 \,\delta^3 (\vec{r}) $. The HF  proper self-energy  reduces to:
$ \Sigma ( \vec{r}_2 - \vec{r}_1)= - g^2 \, a^3 \,G(0,\tau )\delta^3 (\vec{r}_2 - \vec{r}_1) $.
This  implies: $ \tilde{\Sigma}(\vec{p} ,\tau )= - g^2 \, a^3 \,G(0,\tau )$.
 \subsection{ The reduction of the HF  integral equation to a numerical equation. }
 The  Green function in momentum space
is then given by :
 \be
\widetilde{G}(\vec{p},\tau)=\widetilde{g}(p^2,\tau)= \((  h_0(p^2)+ \tau + g^2 \, a^3 \,G(0,\tau ) \))^{-1}.
\label{gptau}
\ee
 By computing $ G(\vec{r}, \tau)$  from the above equation by a Fourier integral and then  taking
the limit $ \vec{r} \to 0 $, one gets a self-consistent equation   giving $ G(0,\tau ) $:
\be
 G(0,\tau) = \frac{1}{ 2 \pi^2} \int_0^\infty\frac{  p^2 d\, p \,}{  h_0(p^2) + \tau + g^2 \, a^3 \,G(0,\tau ) }  \; \; .
\label{hfG}
\ee
From now on, we shall take $ h_0(p^2) =h^{(4)}(p^2)=\frac{1}{6}\, p^2+\frac{1}{180}\,p^4 $.  ( Remember
that our length unit is $ b $). With our choice of  $ h_0(p^2) $ the integral in the r.h.s.  is clearly convergent,
while it would have   been  linearly divergent in the case of the free Gaussian chain.
We have found  convenient to  use  instead of $ G(0,\tau)$    the  following self-consistent function:
\be
  \mu(\tau)= \tau + g^2 \, a^3 \,G(0,\tau ) .
\label{mutau}
\ee
 To solve  exactly  the  HF  equation we have introduced the algebraic function :
\be
 F(\kappa,\tau)=\int_0^{\infty}\frac{p^2}{p^2 + p^4\,{\kappa }^2 + \tau }=\frac{\pi \,\left( -1 + 2\,\kappa
\,{\sqrt{\tau }} +
      {\sqrt{1 - 4\,{\kappa }^2\,\tau }} \right) }{{\sqrt{2}}\,\kappa \,
    {\sqrt{1 - 4\,{\kappa }^2\,\tau }}\,
    {\sqrt{1 - {\sqrt{1 - 4\,{\kappa }^2\,\tau }}}}} \; ,
 \ee
where we have made explicit for clarity  the cutoff parameter $  \kappa ^2= b^2/\Lambda^2 =1/30 $.
 In order to study  the cross-over  occurring     in the vicinity of $ f=0 $ , the following approximation of
 $F(\kappa,\tau)$ is useful:
\be
  F_0(\kappa,\tau)=\frac{\pi }{2\,\kappa } \,\left( 1  - \kappa \,\sqrt{\tau } +
      \frac{3\,{\kappa }^2\,\tau }{2} + O\,{\left( \kappa \,{\sqrt{\tau }} \right) }^  {3}\right) .
\ee
We are now  ready  to write down  the  self-consistent equation  giving  $\mu(\tau) $ :
\bea
    \mu( \tau) & = &\tau+ \frac{\lambda}{6} \, F\((\kappa ,6  \mu( \tau)\)) , \nonumber\\
     \lambda &= & \frac{18}{\pi^2}\, g^2\, a^3 .
\label{selconsismu}
\eea
Remembering that  $\widetilde{g}(-f^2,\tau) =\widetilde{z}(-f^2,\tau)=  \(( h_0(-f^2 )+ \mu(\tau) \))^{-1}$ is the  Laplace
 transform  the searched-for quantity $ \widetilde{Z}( -f^2, N) $, one has to perform  the following   inverse Laplace transform:
$$
\widetilde{Z}(- f^2, N)= \frac{1}{ 2 \, \pi \,i}\int_{ \epsilon-i\,\infty}^{ \epsilon+i\,\infty} \ d \,\tau\exp(N\, \tau)
\(( h_0(-f^2 )+ \mu(\tau) \))^{-1} .
$$
  It appears that the relevant  singularities  of $ \widetilde{g}(-f^2,\tau) $,
 in the complex half plane $ \Im(\tau) <0 $,  are
lying   along  the negative real half axis. As a consequence,  the  integration contour
along the imaginary $ \tau $ axis can be
folded around the negative  real  half  axis. This  leads  to the following integral  formula :
\be
\widetilde{Z}(- f^2, N)= \frac{1}{ 2 \, \pi \,i}( \int_{0+i\,\epsilon}^{ -\infty +i\epsilon} +\int_{-\infty -i\,\epsilon}^{
0 -i\epsilon}) d \,\tau\exp(N\, \tau)
\(( h_0(-f^2 )+ \mu(\tau) \))^{-1} .
\label{invlap}
\ee
\begin{figure}
\centerline{\epsfxsize=120mm  \epsfbox{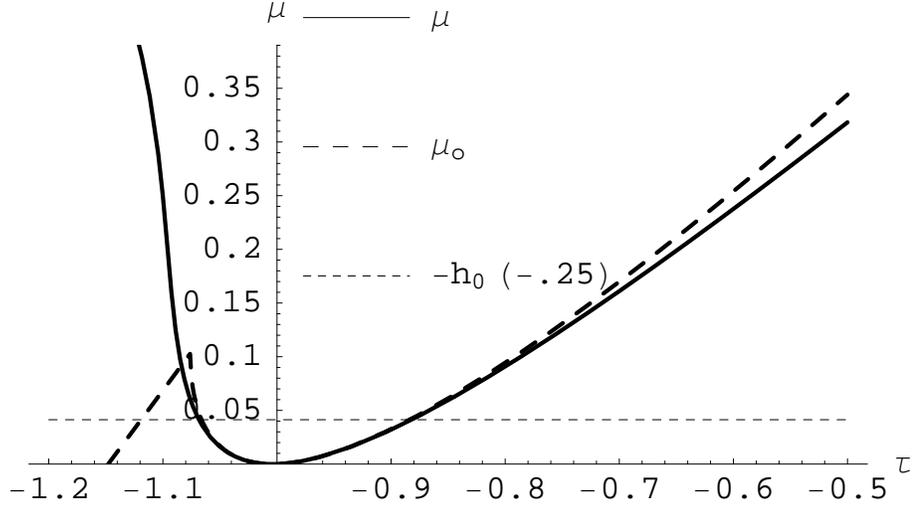}}
\caption{\footnotesize The partition function $\widetilde{Z}(- f^2, N) $ is obtained  from the inverse Laplace transform of
$\(( h_0(-f^2 )+ \mu(\tau) \))^{-1}$. The result is dominated by the  pole residues  of the integral. Since
$h_0(-f^2) \leq 0 $, we have to look for  the positive solutions of   $ \mu( \tau)  = \tau+ \frac{\lambda}{6}
F\((\kappa ,6  \mu( \tau)\)) $. The continuous line  gives  the self consistent function   $\mu( \tau)$  for
$ \lambda=0.7 $.
The dashed line is associated with   the approximate function   $\mu_0( \tau)$,  obtained
with the simplified  function  $ F_0(\kappa ,\tau ) $, valid  for small  $ \tau $  up to corrections  of  order
$(\frac{\tau}{30})^{3/2}$. The figure shows that  this   approximation  is adequate  in  the region below
 the horizontal dotted line, which corresponds to   the low force  interval $ 0 \leq  f \leq 0.5$   . }
\label{fig3}
\end{figure}
\subsection{ Analytical computations of the extension vs force  curves in the low force regime.}
  Let us consider first the low force regime $0 \leq f \leq 0.5 $ {\it i.e }  $ 0 \leq -h_0(-f^2)\leq 0.0413$. We
are clearly  interested  by the value of  $\mu(\tau) \simeq  -h_0(-f^2) \leq 0.0413  $. The use of the approximation
  $F(\kappa,6 \mu(\tau))\simeq  F_0(\kappa, 6\mu(\tau))$  is  justified   here  by
  the small  value expansion parameter :
$ \kappa \sqrt{6 \mu(\tau)} \leq 0.091 $. The  corresponding approximate function  $ \mu_0(\tau) $
differs from the exact one  by less than  $ 0.5 \%$ if $0 \leq f \leq 0.5 $ , as it  is seen clearly on FIG. \ref{fig3}.
Let us look first for  the value $\tau=\tau_c =\frac{-\left( {\sqrt{\frac{5}{6}}}\,\pi \,\lambda  \right) }{2}$
such that $ \mu(\tau_c)=0 $.  It is convenient  to define  the  translated  variable
$\sigma =\tau -\tau_c $. To get an explicit expression of   $ \mu_0(\tau) $, what we have done  first is to
solve the self-consistent  equation for $ \tau  >0   $. In this region ,  $ \mu_0(\tau) > 0 $ , so that the function
$ F_0(\kappa, 6\mu_0(\tau)) $ is regular.  We   perform  next an analytic continuation  toward  the point $\tau=\tau_c$
 where $ \mu_0(\tau) $ vanishes. According  to eq. ( ref{invlap}), we have to follow
    paths  running just above ( and  below ) the  half
negative  half axis. This insures that   $ \sqrt{ \mu^2}= \mu $ in the vicinity of  $ \tau= \tau_c$.
We arrive  then  to the   following explicit  formula, valid in the
region of physical interest:
\newpage
\bea
\mu_0( \tau_c+ \sigma ) &= &
\frac{240\,{\sigma }^2}
  {240\,\sigma + \pi \,\lambda \,\left( 5\,\pi \,\lambda  - 6\,{\sqrt{30}}\,\sigma  +
       {\sqrt{2400\,\sigma + 5\,\pi \,\lambda \,
            \left( 5\,\pi \,\lambda  - 12\,{\sqrt{30}}\,\sigma \right) }} \right) }  \nonumber  \\
 & =& \frac{24\,{\sigma }^2}{{\pi }^2\,{\lambda }^2} +
  \frac{144\,\left( -40 + {\sqrt{30}}\,\pi \,\lambda  \right) \,{\sigma}^3}
   {5\,{\pi }^4\,{\lambda }^4} + O(\sigma ^4  )   \; .
\label{formu0}
\eea

 Some remarks are in order : first, the above  expression   has for  small value of $ \tau-\tau_c $,
  the parabolic shape  exhibited  on FIG.\ref{fig3}, second  a branch  point occurs
 for $ \tau-\tau_c=\tau_b=\frac{5\,{\pi }^2\,{\lambda}^2}{-480 + 12\,{\sqrt{30}}\,\pi \,\lambda }$.
It manifests  its presence  on the curves of FIG.\ref{fig3} by a spike. (Below the
branch point it  is actually  $\Re( \mu_0 ( \tau) ) $ which is plotted.)
Such  a singularity does  not seem to show up  in the range of interest
for the exact function $\mu(\tau)$ and it is  likely
to   be an artefact of  the approximation     $F(\kappa,6 \mu(\tau))\simeq  F_0(\kappa,6 \mu(\tau))$.
Nevertheless,  we have computed   explicitly its contribution to  the inverse
Laplace transform and it  was  found  to be   negligible compared to the dominant poles contributions.
Finally it is of interest to note that the coefficient of the series expansions with respect to $ \sigma$ are divergent
in the weak coupling limit $ \lambda \rightarrow  0$. Furthermore,  we have checked the identity of
 series expansion in power of $ \lambda $ obtained  by two  different ways: one, by a direct expansion of the above
exact expression of $\mu_0 ( \tau) $, two, by solving the self-consistent equation by a perturbation method.
The coefficients  of the various powers of $ \lambda $ have    a branch cut at $ \tau =0 $  which disappear
when the exact summation is performed. The above remarks    provides an illustration, on a simple case,
 of the importance of non-perturbative  effects  in the self-avoiding  polymers problem.

\begin{figure}
\centerline{\epsfxsize=120mm  \epsfbox{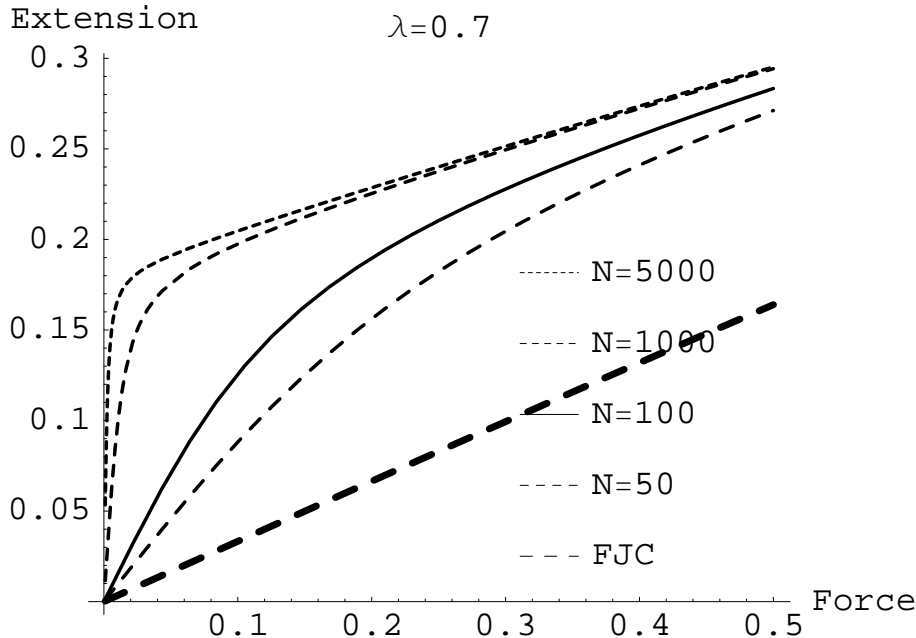}}
\caption{  \footnotesize On this figure  we have displayed a set of extension vs force  curves in the low
energy regime, for typical values of the monomer number $N$.  The bottom thick dashed curve represents
the FJC extension vs force curve with no self-avoidance.
 The most  remarkable feature is  the steady increase with   $N$  of the slopes at the origin ($f=0$).
 We note also that in  the limit of very large $N$, the extension jumps to a finite value at very small force.
This suggests the presence of a cross-over phenomena   near  $ f=0 $. This could be associated
 with the  partition  function Laplace transform  poles crossing  at  the  origin.  }
\label{fig4}
\end{figure}

The two  poles $\tau_{1,2}(f,\lambda) $ of $\(( h_0(-f^2 )+ \mu_0(\tau) \))^{-1} $  are   easily  computed:
 \bea
\tau_{1,2}(f,\lambda)&=&  \tau_c  \pm \frac{\pi \,\eta(f) \,\lambda }{2\,{\sqrt{6}}} +
  \eta (f)^2\,\left( 1 - \frac{{\sqrt{\frac{3}{10}}}\,\pi \,\lambda }{4} \right) \nonumber , \\
\eta(f) &=& f \sqrt{1- \frac{1}{30}  \,f^2 }.
\label{fortaupole}
 \eea
We have  in hands all the information needed to compute the partition function  $\widetilde{Z}(- f^2, N) $
given by eq. (\ref{invlap}). Ignoring the branch point contribution, we are left with the
residue contributions relative to the  two poles at $ \tau= \tau_{1,2}(f,\lambda) $:
\be
\widetilde{Z}(- f^2, N)= \sum _{i=1}^2\exp\(( N\, \tau_{i}(f,\lambda) \))
\(( { \frac{\partial  \, \mu_0\((  \tau_{i}(f,\lambda) \))} {\partial \tau} } \))^{-1} .
 \ee
The relative extension $ \zeta(f,N)= \langle \, z(N) \,\rangle /N $ is then given by the logarithmic
derivative:
\be
\zeta(f,N)= -\frac{1}{N}\, \frac{ \partial}{\partial \,f} \log \(( \widetilde{Z}(- f^2, N) \)) .
\ee
Using the explicit formulas for $ \mu_0(\tau) $ and $\tau_{i}(f,\lambda)  $ given in eq. (\ref{formu0})
and (\ref{fortaupole})$, \zeta(f,N) $ can be computed analytically. The corresponding expression is rather
involved and not very illuminating. A better illustration of what is going on in the low force regime
is obtained by looking at FIG.\ref{fig4} where  relative extension vs force curves
 $ \zeta(f,N)= \langle\, z(N) \,\rangle /N$ are displayed for
increasing values  of
  the total  chain monomer  number N.
\subsection{Comparison  of HF extension slopes at the origin (f=0) with the results  derived from
RG arguments.}
The most salient feature of these curves lies in the fact that  the extension vs force  curve slope,
 $ s( N,\lambda )= \frac{\partial E(f,N)}{ \partial \,f} {\vert}_{f=0} $,  increases
with $N$, while it stays constant in the case of the FJC model. This   kind of behaviour
 is found under a less   pronounced form in the Edwards model (Gaussian chain with short range
 self-avoidance), using renormalization group arguments (RG). We would like to stress that a quantitative
comparison of the two approaches  is not a simple problem. We have noted previously that the model
used in the present paper can be viewed as  a version of the Edwards model, endowed with a
a  {\it fixed cutoff }$ \Lambda^2 =\frac{30}{b^2}$. It is then clear that  under such circumstances
 RG arguments are not appropriate.

    \begin{figure}
\centerline{\epsfxsize=100mm  \epsfbox{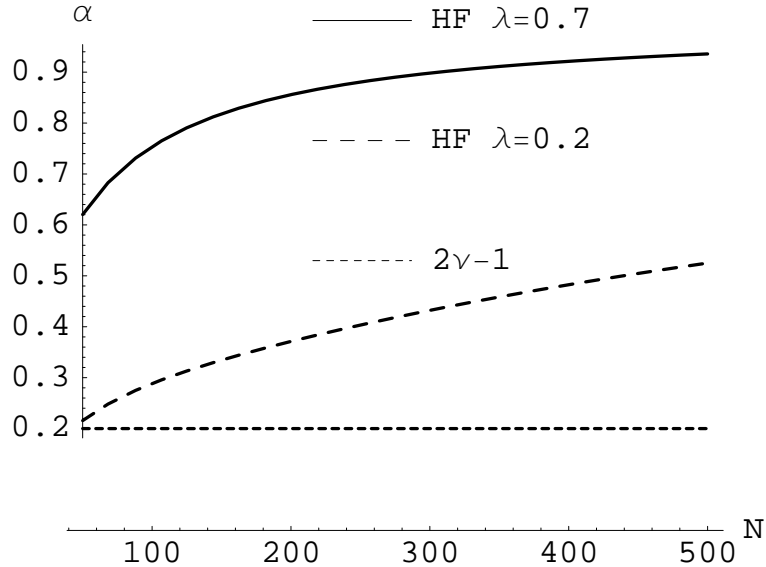}}
\caption{ \footnotesize  In this figure we present the results of a simple analysis of the  variation  of the slope of the
extension vs force curve $s( N,\lambda )$  at $ f=0 $   with the monomer  number $ N$. We have plotted
the characteristic quantity  $ \alpha(N,\lambda) $ defined as $ N $ times the logarithmic
derivative  of the slope  at the origin with respect to $ N $.
 If    $ \alpha(N,\lambda) $ stays constant,   the  slope behave as  $ N^{\alpha }$
or, in other words, it obeys a scaling law. RG arguments applied to the Edwards model gives
$\alpha = 2 \, \nu -1 $, where $\nu  \simeq  0.6  $ is the Flory  coefficient. Our model exhibits  a
moderate but significant variation of $\alpha(N,\lambda) $   with $ N $.
Moreover,  $\alpha(N,\lambda) $ is not "universal ", in
 the sense that it is roughly proportional to the coupling  constant $ \lambda  $. We would like to
stress that this difference was to be expected. As we have pointed out  previously,   our model can be
viewed as a fixed cutoff ( $ \Lambda ^2 = 30 /b^2 $)  version of the Edwards  model and,
as a consequence, RG arguments  are not valid. }
\label{fig4bis}
\end{figure}

 For exhibiting  the   difference of behavior  in the vicinity
of $ f=0 $,  it  is convenient to introduce the following quantity :
  $ \alpha(N,\lambda)= N \frac{\partial}{ \partial N}\log\(( s( N,\lambda ) \))$.
Its  physical interpretation becomes  more  apparent by noting that if $ \alpha(N,\lambda) $ is constant,
 then the slope $  s( N,\lambda ) $ is proportional to $ N^{\alpha}$. This corresponds to
 the scaling   prediction \cite{PGDG} for the
Edwards model with $\alpha = 2 \nu -1$, where $\nu \simeq \frac {3}{5} $ is the
Flory coefficient,
 which  governs  the variation of the  root mean square radius of the Edwards chain in absence of
stretching force : $ \sqrt{ \langle \, r ^2(N) \, \rangle }\propto N^{\nu} $.
Curves giving  the variation of  $ \alpha(N,\lambda) $ are displayed on FIG.\ref{fig4bis}. In the case
of  the present  HF computations,  there is, for  fixed $ \lambda $,
a  moderate but significant increase of $ \alpha(N,\lambda) $ with $ N $.
 Moreover, for fixed  $ N $,  when $\lambda$  goes
 from  $0.2 $  to  $0.7 $,   $ \alpha(N,\lambda) $ is multiplied by  a factor varying from $3$ to $2$
when N increases from $ 50$ to $ 500$.

 These features  are likely to be  associated with the fact that we are dealing
with a fixed cutoff  field theory problem  but  they could also
 be partly due to an artefact of the Hartree-Fock approximation.
In order to check this latter point,  one should  try  to estimate the corrections to the HF approximation.
In the particular case of the  short range repulsive  potential  $  g^2 V(\vec{r})= a^3 g^2 \delta(\vec{r}) $,
  the proper self-energy  correction  $\Sigma^{(2)} ( \vec{r}_2 - \vec{r}_1) $
 is given by  the rather simple expression:
\be
\Sigma^{(2)} ( \vec{r}_2 - \vec{r}_1)=g^4 a^6 \((  G( \vec{r}_2 -\vec{r}_1) \))^2
\ee
The  self-consistent   integral equation cannot be reduced anymore to the solving of   a numerical equation.
    The resulting  non-linear integral equation  can be solved by an iteration procedure starting from the
HF approximation.  Such a calculation,  which looks feasible, is clearly  of considerable interest  but
it falls outside the scope of the present exploratory  paper.
\section{ Comparison of the Hartree-Fock  Force  vs Extension Curves   Computations   with  the Results of
Monte-Carlo Simulations }
In this last section we would like to compare the results
of the Hartree-Fock method applied to  the self-avoiding effects  for a continuous version of the FJC model
with those  obtained  by a Monte-Carlo simulation of  a self-avoiding freely joining chain.

\begin{figure}
\centerline{\epsfxsize=120mm \epsfbox{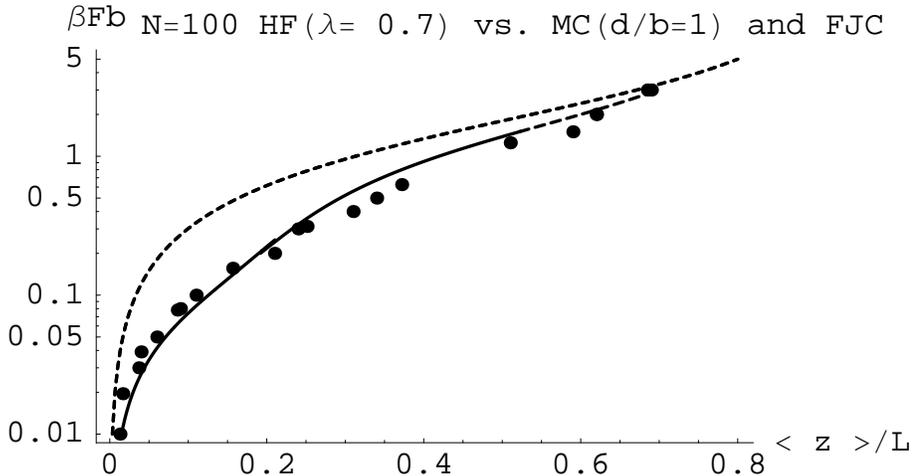}  }
\caption{ \footnotesize This figure presents a comparison  of the  force vs extension curves for $ N=100 $ obtained, on one hand,
from the HF analytic computations of the present paper, on the other hand, from MC simulations
performed by the ENS experimental  group
 Our analytic treatment, represented by the continuous curve, is strictly valid
in the low to medium force range $ 0 \leq f \leq 1.5 $. Nevertheless, we have performed
 a reasonably safe extrapolation -  the dotted branch of the  curve- in  order to cover
the higher  force range $ 1.5 \leq f\leq 3 $. The choice of the value of our single parameter: $ \lambda = 0.7 $
does  not  come out from a least square   fit but  from a trial and error procedure.  In the MC simulations
  self-avoidance was implemented by  keeping only  the  configurations such  that the distances
between any  two non -adjacent monomers  are  larger  than a fixed length $d$. In the
actual simulations $ b$ was taken equal to
the fixed distance  $b$  between   adjacent polymers. The fairly good agreement between
these  two different   approaches is quite remarkable and it may suggest the shape of the force curves is
not very sensitive to the range of the repulsive potential if $ d\leq b $.  }
\label{fig6}
\end{figure}

We begin by saying some words about the way we have obtained the theoretical
force vs extension curve  displayed  on  FIG.\ref{fig6}.
In the range: $ 0 < f < 0.25 $ we have used  the analytic formulas  of the previous section, which are valid in the
low force  regime. In the medium  force interval : $  0.2   < f  <1.5 $ the computation  was  performed  with the
exact solution  $ \mu(\tau )$  of the self-consistent equation (\ref{selconsismu}). A considerable simplification was achieved
by  noting  that,  if  $ f > 0.2 $,  the pole residue  at $ \tau=\tau_{1}(f,\lambda) >\tau_{2}(f,\lambda) $
gives the   dominant contribution,  by more than two orders of magnitude.

Above $ f=1.5 $ we are leaving the domain
of strict  validity of our computation. We have  performed an extrapolation  to cover the interval  $ 1.5 < f  < 3.0 $.
It corresponds to the dashed part of the theoretical  curve of FIG.\ref{fig6}.
We do not extrapolate directly  the final  result but rather   the quantity
  $\widetilde{g}(-f^2,\tau)^{-1} = h_0(-f^2 ) + \mu(\tau)$.
 For the first  term, we set $ h_0(-f^2 )= h_0^r(-f^2) $,  instead of keeping   $ h_0^{4} (-f^2)$
 which  would lead  to an inadequate  force vs  extension  curve for $ f > 2$, as it appears  clearly on FIG.\ref{fig1}.
  Concerning $\mu(\tau )$,  we note two things: first,
  the replacement   $ h_0^{4} (p^2) \rightarrow h_0^r(p¨^2) $
will only affect the high $p$ tail of the  convergent integral   giving the function
$  F(\kappa, \tau ) $ (see FIG.\ref{fig2}), second, the non corrected   $\mu(\tau )$  varies linearly
in the interval of interest:  $ 0.35 <  \mu(\tau )  <1.2 $.  It look then
reasonable to perform our extrapolation  by keeping  the self-consistent function $\mu(\tau)$ given
by eq.(\ref{selconsismu}).

The M.C simulation results displayed   as big dots on FIG.\ref{fig6}  have been obtained
by the authors of reference \cite{mndessin},  with   a chain of
 $N=100$ monomers. Two adjacent monomers are freely connected by a segment of length $b$.
The repulsive potential between two non-adjacent monomers  is taken to be of the square wall type:
$ V( \vert \vec{r}_1 - \vec{r}_2  \vert)=V_0 $ if $\vert \vec{r}_1 - \vec{r}_2  \vert) \leq  d $ and 0
otherwise. In the simulation, it is tacitly assumed that $V_0 \gg k_B T$ . In practice, this means that the only
accepted configurations are such that the distances between  any pair of two non- adjacent  monomers are
$\ge d $. The short range  potential   used in our calculation corresponds to a different limit :
$ V_0 / k_B T \rightarrow \infty \; ,\; d \rightarrow 0$, with $ \frac{ 4\pi } { 3} d^3 V_0/k_B T $ going to
a finite limit  $ \gamma b^3 $. The limiting potential $ V( \vert \vec{r}_1 - \vec{r}_2  \vert) $  reduces
to  the zero range potential $ \gamma b^3 \delta( \vec{r}_1 - \vec{r}_2 )$. The coupling constant $\lambda$ used
in our computation is to be identified with $ \frac{18}{ \pi^2 } \gamma b^3  $.
Despite the  difference of the
 two kinds of limit, we see on FIG.\ref{fig6}  that the M.C. simulation results with $ d=b $
 are fairly well fitted by our  HF  curve,  if we choose     $\lambda =0.7$.
( We  work  in a  system where $b $ and $ k_BT$ are taken  respectively as units of length and energy).
 In physical terms, this may suggest  that the shape of the  force vs extension curve is  rather insensitive to the
 potential range $d$,  as long as it is equal or smaller  than  the monomer length $b$.  This important  point
 could be  confirmed in two ways: one  by MC simulations  with $ d/b<1 $, second by solving numerically
 the HF integral equation (\ref{hfinteq}), involving  finite range  Yukawa or exponential potentials. If  this last
 numerical   program  can be achieved  successfully, it will also  open the road to  HF method applications
 to biological  polymer like ssDNA, using  more  realistic monomer-monomer  potentials.

 \section*{Conclusion}
 The exploratory investigation of the HF method, as a tool for  the study of flexible
 self-avoiding polymers,  appears encouraging enough to justify    an extension  of the present work
 in several directions. The most urgent task is perhaps to get an estimate of the "non planar" diagrams
 contributions to the self-consistent  equation, along the lines suggested at the end of section 3.3.
 Despite the good agreement with the  MC simulations, it is far from obvious that the
 main physical results of the present  work stay robust vis-\` {a}-vis such corrections.

The second  road to be explored is the numerical solving  of the  self-consistent  equation (\ref{hfinteq})
  with  more realistic monomer-monomer   potential  $ V(r)$, say a Yukawa potential,  rather than
the short range limit used in the present paper.
The numerical work load will not be sensibly increased if one takes  for $ V(r)$,
a superposition of     Yukawa and exponential potentials, adjusted in such a way that
the  monomer-monomer interactions  are attractive for distance about the hydrogen bound length
 and  become repulsive when one reaches distances in the range of the Debye length.
 The MC simulations and the HF model  force vs extension curves   deviate from
 the data relative to  inhibited hydrogen bonding \cite{mndessin},  in the low force regime.
 In particular the experimental curves suggest that the molecular
 chain extension vanishes for a small  but finite
 value of the stretching force, about $ 0.05 pN$. This could be an indication
 that  hairpin structures are still present at very low force, say below $ 0.1 pN$.
 It will be of interest to see, if by adjusting the strength of the attractive monomer-monomer
 interactions, our HF model is able to reproduce the data in  the whole range of force.

\section*{Acknowledgments}
It is a great pleasure to thank the  members of the biomolecular physics group
at ENS, M.-N. Dessinges, B. Maier, M. Peliti, D. Bensimon and V. Croquette,
for communicating their results before publication. We are very grateful to D. Bensimon  for
his  careful reading of the   manuscript and for many judicious remarks.

 \end{document}